\documentstyle[preprint,prl,aps]{revtex}

\newcommand{\tj}{$t$-$J$}
\newcommand{\dxy}{$d_{x^2-y^2}$}
\newcommand{\etal}{{\it et al}}
\newcommand{\ij}{\langle ij\rangle}
\renewcommand{\S}{{\vec S}}  
\newcommand{\grapprox}{\stackrel{>}{_\sim}}
\newcommand{\lapprox}{\stackrel{<}{_\sim}}

\begin{document}
\draft

\title{Ground State Properties of the Doped 3-Leg $t$--$J$ Ladder}
\author{ Steven R.\ White$^1$ and D.J.\ Scalapino$^2$}
\address{ 
$^1$Department of Physics and Astronomy,
University of California,
Irvine, CA 92697
}
\address{ 
$^2$Department of Physics,
University of California,
Santa Barbara, CA 93106
}
\date{\today}
\maketitle
\begin{abstract}
\noindent 

Results for a doped 3-leg \tj\ ladder obtained using the
density matrix renormalization group are reported.  
At low hole doping, the holes form a dilute gas with a
uniform density. The momentum occupation of the odd band shows a
sharp decrease at a large value of $k_F$ similar to the behavior
of a lightly doped \tj\ chain, while the even modes appear
gapped. The spin-spin correlations decay as a power law
consistent with the absence of a spin gap, but the pair field
correlations are negligible. At larger doping we find evidence
for a spin gap and as $x$ increases further we find 3-hole
diagonal domain walls. In this regime there are pair field
correlations and the internal pair orbital has \dxy-like
symmetry. However, the pair field correlations appear to fall
exponentially at large distances.

\end{abstract}

\section{Introduction}
The recently discovered $n$-leg ladder cuprate
materials\cite{ishida} form an
interesting testing ground for ideas regarding strongly correlated
electron systems.
Just as for the two-dimensional layered cuprates, the \tj\ Hamiltonian
is believed to provide a basic model which contains the general features
of the ladder systems.
In particular, the undoped, even-leg Heisenberg ladders have been shown
to exhibit a spin gap\cite{riceone,rvbprl}, 
while the odd-leg ladders have no spin gap, in
agreement with experiment\cite{azuma}.
Various calculations on the doped 2-leg ladder find that in the ground
state the doped holes form \dxy-like pairs, and the system is
characterized by power-law pair-field and $4\,k_F$-CDW 
correlations\cite{noack,troyer,balents}.
A recent study of the 4-leg doped \tj\ model\cite{fourchain} 
found evidence for three
types of phases, depending upon the ratio of $J/t$ and the hole doping
$x$.
At low doping, when holes are first added to the insulating state, they
form a dilute gas of \dxy-like pairs.
At higher doping, the holes arrange themselves into fluctuating domain
walls while maintaining significant \dxy\  pair-field correlations.
Finally, at sufficiently large $J/t$ values ($\grapprox 1.5$), phase
separation occurs.
Based upon these results, it is interesting to explore what happens when
an odd-leg ladder is doped.
Here we discuss results obtained from density matrix renormalization
group (DMRG) calculations\cite{dmrg,cavo} for a doped 3-leg \tj\ ladder.

The \tj\ Hamiltonian which we will study has the form
\begin{equation}
{\cal H} = J \sum_{\ij} \left(\S_i\cdot\S_j - {\textstyle{1\over4}}\,
n_in_j \right) 
- t \sum_{\ij,s} P_G \left( c^\dagger_{i,s}c_{js} +
c^\dagger_{j,s}c_{i,s} \right) P_G.  
\end{equation}
Here $\ij$ denotes nearest-neighbor sites on the ladder, $s$ is a spin
index, $\S_i$ and $c^\dagger_{i,s}$ are electron spin and creation
operators, $n_i = c^\dagger_{i,\uparrow}c_{i,\uparrow} +
c^\dagger_{i,\downarrow}c_{i,\downarrow}$ and the Gutzwiller projector
$P_G$ excludes configurations with doubly occupied sites.
For the DMRG calculations discussed here we have used open boundary 
conditions for $L\times3$ clusters and have set $J/t=0.35$, which is
close to the value expected for the cuprates.
From 800 to 1500 states were kept per block. Truncation errors
were typically $\sim 10^{-5}$.

In the next section, II, we begin by showing the most probable ground
state hole configurations and the over-all structure of the charge and
spin correlations for different hole densities.
While these projections show only a caricature of the groundstate which
contains a huge superposition of states corresponding to large
fluctuations of the structures shown, they provide a useful picture for
the discussion which follows.
In particular, the average hole rung density shows that the domain wall
structure seen in the projections can survive in a ground state
expectation value.

In Section III, we calculate the momentum distribution of the holes and
use it to discuss the nature of the quasiparticle excitation bands.
Following this, in Section IV, we examine spin correlation and the
behavior of the spin gap in the doped system.
In Section V we discuss the orbital structure of the
pairs and the pairing correlations.
Section VI contains our conclusions.

\section{Charge Density and Spin Structure}

As we have previously discussed\cite{holestructures}, 
two holes doped into a 3-leg 
\tj\ ladder are not bound for physically relevant values of $J/t$.
However, we have found that as holes are added at a fixed value of
$J/t$, the system evolves from a gas of holes to an array of domain
walls as shown in Fig. 1.
This figure shows the most probable configuration of holes in the system
for $J/t = 0.35$ obtained by maximizing the ground state expectation
value of
\begin{equation}
P(\ell_1,\ell_2,\ldots) = \prod_{i=1} p(\ell_i), 
\end{equation}
with $p(\ell_i) = (1-n_{\ell\uparrow})(1-n_{\ell\downarrow})$ the hole
projection operator for the $\ell^{\rm th}$ lattice site.
Here $(\ell_1,\ldots,\ell_{N-1})$ are chosen to give the most probable
location of $N-1$ holes, and for this configuration, the diameter of the
dark circles gives the probability of finding the $N^{\rm th}$ hole on a
given site.

At low hole concentrations, the most likely configuration consists of
individual holes as shown in Fig. 1(a).
The most probable location for the holes are on the outer legs
and as we will see reflect the fact that in the dilute limit the
holes are doped into the odd single particle band.
As the density $x$ of holes increases, fluctuating domain wall-like
arrays appear.
This initially occurs as a domain wall running down the center
chain as seen in Fig. 1(b). At higher densities, fluctuating
diagonal three-hole domain walls appear as shown in Fig. 1(c).

The local structure of these diagonal domain
walls is similar to that of the diagonal domain walls observed on 4-leg
ladders\cite{fourchain}.
Figure 2 shows a section of the lattice which contains a three-hole
domain wall.
Figure 2(a) shows the strength of the exchange field
$-\langle\S_i\cdot\S_j\rangle$ when the holes occupy their most likely
positions.
The strong diagonal singlet correlations in Fig. 2(a) are similar to
those which are found for \dxy-pairs on the even leg ladders as
well as on 2D lattices.
These diagonal singlet correlations reflect the fluctuating nature of
the wall which reduces the kinetic energy of localization, while
at the same time minimizing the exchange energy.
In Fig. 2(b), an external staggered magnetic field has been applied to
the left-hand end of the ladder, and one can see that the
antiferromagnetic spin background undergoes a $\pi$-phase shift as it
crosses the domain wall.

Another view of this is given in Fig. 3(a) which shows the
average charge density and the spin structure. Here, as in Fig.
2(b), a small staggered magnetic field has been applied to the
left hand end. Fig. 3(b) shows a view of the longitudinal
domain wall of Fig. 1(b). Here there is a small staggered field
along the bottom leg. The resulting spin moments on the top leg
are $\pi$-phase-shifted with respect to the bottom leg.
The crossover from longitudinal to transverse (diagonal) domain
walls appears to be smooth as $x$ increases.

The development of transverse domain walls is also 
clearly evident in Fig. 4, which shows 
the average rung density
\begin{equation}
\langle n_\ell\rangle = {\textstyle{1\over3}} \sum^3_{j=1} 
\langle n_{\ell j}\rangle.
\end{equation}
The open boundary conditions break
translational invariance, allowing density variations to be
seen.
At low hole and moderate hole densities (the lower two curves), 
the average rung density is fairly uniform, corresponding
to individual holes and to a longitudinal domain wall.
At a filling of 18 holes on a $32\times3$ ladder, corresponding to
$x=0.1875$, six three-hole diagonal domain walls are clearly evident in the top
curve in Fig.~4.

\section{Momentum Distribution of the Holes}

The one-electron eigen operators of a non-interacting 3-leg ladder have
the structure
\begin{equation}
\gamma^\dagger_{k_x,k_y,s} = \sum^3_{\ell=1} {\sin k_y\ell \over \sqrt2}\,
c^\dagger_{k_x,\ell,s}, 
\end{equation}
with $k_y = \pi/4$, $\pi/2$, and $3\,\pi/4$.
The corresponding eigen energies are
\begin{equation}
\varepsilon_k = -2\,t(\cos k_x + \cos k_y). 
\end{equation}
The states $k_y = \pi/4$ and $3\,\pi/4$ are even under reflection about
the center chain, while the $k_y = \pi/2$ state is odd.

Using the DMRG technique, we have calculated the equal time expectation
value $\langle\psi_0| c^\dagger_{i_x,i_y,s} c_{j_x,j_y,s} |\psi_0\rangle$
and from this constructed the momentum occupation expectation values for
the three $k_y$-bands
\begin{equation}
\langle \psi_0| n_{k_x,k_y,s} |\psi_0\rangle =
\langle \psi_0| \gamma^\dagger_{k_x,k_y,s} \gamma_{k_x,k_y,s}
|\psi_0\rangle. 
\end{equation}
Figure 5(a) shows the momentum occupation for the three bands at low
doping, $x = 0.042$.
For comparison, the momentum occupation for a single chain \tj\ model
with $J/t=0.35$ and a hole density $x=0.1$ is shown in Fig. 5(b).
The decrease of $n(k)$ sharpens as the length of the \tj\ chain is
increased and marks the Fermi wave vector $k_F$ of the single-chain
Luttinger liquid.
The structure of $\langle n_{k_xk_y}\rangle$ for the odd $k_y=\pi/2$
band is similar to that of the single chain \tj\ system. 
This suggests that at low doping, the $k_y=\pi/2$ band of the 3-leg
system is not gapped at the Fermi surface, while the two even bands at
$k_y=\pi/4$ and $3\,\pi/4$ appear to be gapped.
When the density of holes is increased, all three bands appear to be
gapped, as shown in Fig. 6(a) and (b) for $x=0.125$ and $x=0.1875$
respectively.
These conclusions are confirmed by direct observation of 
$\langle\psi_0| c^\dagger_{i_x,i_y,s} c_{j_x,j_y,s} |\psi_0\rangle$
as a function of $i_x-j_x$ (not shown): at low densities, power
law decay is observed in the odd mode, and exponential decay in
the even modes, while at higher densities, all modes show
exponential decay.

\section{Spin Correlations}

Based upon the quasi-particle momentum distributions discussed in
Section III, we would expect that the lightly doped 3-leg ladder would
exhibit power-law antiferromagnetic correlations arising from the
$k_y=\pi/2$ quasi-particle band.
A log-log plot of $\langle S^z_iS^z_j\rangle$ for $x=0.042$ is shown in
Fig. 7(a) and is consistent with the power law decay one would
expect for a one-dimensional Luttinger liquid.
At higher hole densities $x=0.125$ and 0.1875, the spin-spin
correlations $\langle S^z_i S^z_j\rangle$ are found to decay
exponentially as shown in Fig. 7(b).
This is consistent with the behavior of $\langle n_k\rangle$ discussed
in the previous section.

We have also calculated the spin gap
\begin{equation}
\Delta_S = E_0(S_z=1) - E_0(S_z=0) 
\end{equation}
as a function of the hole doping $x$ for a $44\times3$ ladder.
Here $E_0(S_z)$ is the ground state energy with a given value of total
spin $S_z$.
The result is shown in Fig. 8.
We believe that the non-vanishing spin gap for $x\lapprox 0.05$ is a finite
size effect and that the spin gap will extrapolate to zero in the
low-doping region.
At higher doping, there is a spin gap consistent with the exponential
decay of the spin-spin correlations.

\section{Pairing Correlations}

In order to determine the orbital structure of the pairs, we have
calculated the off-diagonal expectation value
\begin{equation}
_{_{N-2}}\langle \psi_0| c_{\ell\uparrow} c_{n\downarrow}
|\psi_0\rangle_{_N} 
\end{equation}
on a $16\times3$ ladder with $N=4$ holes.
The results of this calculation are shown in Fig. 9(a).
Here, one member of a singlet pair is located at the site marked by the
solid circle.
The shaded circles indicate the amplitude and sign for finding the
second member.
The internal structure of the pair has a \dxy-like form,
although it is somewhat asymmetric with an admixture of $s$ due to
the 3-leg nature of the cluster.

We have also calculated
the pair field--pair field correlation function
\begin{equation}
D(\ell_x) = \langle\psi_0| \Delta_{i_x+\ell_x} \Delta^\dagger_{i_x}
|\psi_0\rangle. 
\end{equation}
Here $\Delta^\dagger_{i_x}$ creates a \dxy-like pair around the $i_x$
site of the middle leg,
\begin{equation}
\Delta^\dagger_{i_x} = c^\dagger_{i_x,2\uparrow} \left(
c^\dagger_{i_x+1,2\downarrow} - c^\dagger_{i_x,3\downarrow} +
c^\dagger_{i_x-1,2\downarrow} - c^\dagger_{i_x,1\downarrow} \right)
- (\uparrow ~~\leftrightarrow~~ \downarrow)
\end{equation}
$D(\ell_x)$ is plotted in Fig. 9(b) for various values of the hole doping.
In the regime of low doping $x \lapprox  .05$, the pair field correlations
are negligible.
At larger values of doping $x$, short range 
pair field correlations are present.
However, as shown by the semi-log plot in Fig. 9(c),
these pairing correlations appear to decay exponentially at
large distances.

While the pair-field correlations do
not exhibit the power law decay found for a two-leg ladder,
there are clearly significant short range pairing correlations
as seen in Fig. 9(b). In order to examine these, we have
calculated the response of the system to a proximity
pairing field
\begin{equation}
H_1 = d \sum_i \left( \Delta^\dagger_{i,i+\hat y} + \Delta_{i,i+\hat y}
\right) . 
\end{equation}
Here
\begin{equation}
\Delta^\dagger_{i,i+y} = c^\dagger_{i,\uparrow} c^\dagger_{i+\hat
y,\downarrow} -  c^\dagger_{i,\downarrow} c^\dagger_{i+\hat y,\uparrow}
\end{equation}
adds a singlet electron pair to sites $i$ and $i+\hat y$.
In this case, the DMRG calculation breaks up the Hilbert space into $N$
modulo 2 sectors, keeping total $S_z$ as a good quantum number.
The response was then determined by measuring the strength of the
induced pair field in both the $\hat x$ and $\hat y$ directions,
$\langle \Delta_{i,i+\hat x}\rangle$ and $\langle \Delta_{i,i+\hat y}
\rangle$ for all sites $i$.
The interaction $H_1$ couples equally to extended $s$-pairs and \dxy\ 
pairs. However, in all the cases that we have studied which show significant
pairing correlations, 
$\langle \Delta_{i,i+\hat x}\rangle$ and 
$\langle \Delta_{i,i+\hat y} \rangle$ have different signs, 
reflecting the \dxy-like symmetry of the response.

In order to compare the \dxy\  response of the 3-leg ladder with that of
the 2- and 4-leg ladders, we have made similar measurements on each of
these ladders.
Figure 10 shows a plot of the average \dxy\  pair field response
\begin{equation}
\langle\Delta_d\rangle = {\textstyle{1\over N}} \sum_i
\left( \langle\Delta_{i,i+y}\rangle - \langle\Delta_{i,i+x}\rangle
\right) 
\end{equation}
for $n=1$-, 2-, 3-, and 4-leg ladders versus doping $x$.
It is clear from this result that the 3-leg ladder has a comparable
\dxy\  pair field response to that of both the 2- and 4-leg ladders.
This is expected from both weak coupling RPA
calculations\cite{bulut} and
renormalization-group studies\cite{arrigoni,kimura,lin}.
However, in the presence of the 3-hole striped domain wall structure
shown in Figs. 1, 2, and 3, it may seem unusual.
In order to understand it in the present framework, 
we have studied the typical hole configurations which
contribute to the pair field correlations for a $12\times3$ ladder with
6 holes\cite{fourchain}.
In Fig. 11(a), we show typical hole configurations in a system
with two diagonal domain walls. These configurations show the
large fluctuations present in the domain walls. These large
fluctuations allow a significant pairing response despite the
presence of the domain wall charge density wave.
In Fig. 11(b), we show some of the specific hole configurations
which give rise to pairing correlations. In particular, we
measure
\begin{equation}
\langle\psi_0| \Delta_{i,i+y} \Delta^\dagger_{j,j+y} 
P(\ell_1,\ell_2,\ldots,\ell_{N-2})
|\psi_0\rangle, 
\end{equation}
where $P$ is given in Eq. (2), and $N$ is the number of holes.
Here the two shaded holes on the right indicate where a singlet pair of
holes is removed and the two shaded holes on the left where a
singlet pair is added. These points are kept fixed.
The black points show typical locations of the remaining four
holes, obtained using a Monte Carlo procedure using DMRG
to measure the probability of a configuration, given by the
absolute value of Eq. (14)\cite{fourchain}.
The configurations show that groups of one, two, and three holes
are common. Most often, a pair is created or destroyed next to
a third hole, thus converting a domain wall into a single
hole and vice-versa.
From these one obtains a general idea of how 
pairing correlations and fluctuating domain walls coexist.

\section{Conclusions}

From these DMRG results for the 3-leg \tj\ ladder, the following picture
emerges.
Initially, when a low concentration of holes is added to the 3-leg
ladder, the holes form a dilute gas with a uniform density, except near
the ends of the open ladder.
The holes tend to occupy the outer legs, associated with the odd quasiparticle mode. 
The momentum occupation
$\langle n_{k_xk_y}\rangle$ indicates that the $k_y=\pi/2$ odd mode is
gapless, while the $k_y = \pi/4$ and $3\,\pi/4$ even modes are gapped.
In this low doping regime, the spin-spin correlations exhibit an
approximate power law decay, and we believe that the spin gap
extrapolates to zero.
The pairing correlations are negligible.

At higher hole densities, fluctuating domain walls appear.
These walls have a similar structure to the domain walls found in
previous DMRG studies.
For $x=0.125$ and 0.1875, the change in momentum occupation $\langle
n_{k_xk_y}\rangle$ is broad for all three bands, consistent with the
finite spin gap observed on the $44\times3$ lattice for 
$x \grapprox 0.06$.
In this regime the spin-spin correlations decay exponentially
and there are significant short-range \dxy-like pairing correlations.

These results have a number of similarities to other recently
reported results for the 3-leg ladder. Rice, et.
al.\cite{ricethreeleg} have
carried out Lanczos calculations on doped $8\times3$ clusters
as well as a mean field analysis. They suggest that below a
critical doping the odd band forms a Luttinger liquid and the
two even bands an insulating spin liquid. Our results for the
momentum occupation $\langle n_{k_xk_y}\rangle$, and the
spin-spin correlations support this picture. 

Renormalization group and bosonization
calculations\cite{arrigoni,kimura,lin} for a 3-leg
ladder suggest that the doped isotropic \tj\  system has a C2S1
phase with two gapless charge modes and one gapless spin mode.
This is also consistent with what we have found. However, in
addition to power law spin-spin correlations, and a gapless odd
mode, this phase is expected to have power law pairing
correlations. However, at low doping we find that pairing
correlations are not present (see Fig. 9(b)). Similarly, 
mean field theory\cite{ricethreeleg} and
RPA calculations\cite{bulut} suggest that the 
system will exhibit pairing correlations at low doping.
However, it appears from the DMRG results that 
a critical hole density of order $x \approx 0.06$ is required before 
this occurs. Rice, et. al.\cite{ricethreeleg} have argued that
this critical doping $x_c$ is associated with a doping
concentration at which the chemical potential becomes equal to
the two hole bound state energy of the spin liquid. From small
cluster Lanczos calculations they estimate $x_c \approx 0.13$, 
but note that it
could be less, while as discussed in Sec. IV, we find an onset of
a spin gap at about half this value.

At larger values of doping we indeed find measurable \dxy-like
pair field correlations. However they appear to decay
exponentially rather than as a power law. Nevertheless, other
measurements, such as the ``proximity'' response of the 3-leg
ladder indicate that significant pair correlations are
present. In this same doping regime we find a spin gap and
observe that the momentum occupation is broadened for all three
quasi-particle modes. Except for the absence of power law pair
field correlations and the prominence of the domain wall
structures, this higher doping regime has a number of features
similar to a Luther-Emery liquid. Rice, et.
al.\cite{ricethreeleg} have suggested that the system makes a transition
into a Luther-Emery liquid when the doping is such that holes can enter the
even-parity quasi-particle modes.

The present work also shows that there is a close relationship between
the appearance of fluctuating domain wall configurations and
pairing\cite{kivelson}. In particular, both longitudinal 
[Fig.  1(b)] and transverse (diagonal) [Fig. 1(c)] domain walls
occur in the doping regime where the pairing correlations
appear. However, in this case, these domains are not produced by
a competition between phase separation and long range Coulomb
interaction\cite{kivelson}. For 3-leg ladders, phase separation
requires unphysically large values of $J/t$ and there is no long
range Coulomb interaction in our model. In the present work, the
domain walls arise as a compromise to minimize the kinetic
energy of the holes and the exchange energy of the spin
background. The local structure of the domain walls shown in
Fig. 2 exhibits hole-hole correlations and exchange bonding
$\langle\S_i\cdot\S_j\rangle$ correlations similar to those
associated with pairs\cite{holestructures}. In addition, as
shown in Fig. 11, pairs can fluctuate between the domain walls.
This is similar to the 4-leg ladder except that in the 3-leg
ladder, the three-hole diagonal walls contain an extra
quasiparticle.

Thus we believe that the common feature associated with pairing
in the $n$-leg \tj\ ladders is the formation of domain walls
containing \dxy\ correlated pairs of holes. On the two-leg
ladder, these domain walls appear simply as rung pairs.
On the 4-leg ladder at a doping where the proximity effect
response $\langle\Delta_d\rangle$ shown in Fig. 11 is large,
one clearly sees the formation of four-hole domain walls
composed of fluctuating \dxy\ pairs\cite{fourchain}. 
In the present 3-leg
system, the domain wall structures form at higher doping levels
and it is in this doping regime that pairing correlations
appear.


We would like to thank L. Balents, M.P.A. Fisher, S.A. Kivelson,
and T.M. Rice for useful discussions.
SRW acknowledges support from the NSF under 
Grant No. DMR-9509945, and DJS acknowledges support from the
NSF under grant numbers PHY-9407194 and DMR-9527304.

\newpage
\begin{figure}
\caption{Maximum likelihood hole configurations at various
densities. In each case, the diameter of the black dots shows
the probability of finding the last hole in the system at each site 
when all the other holes have been projected out in their most likely
configuration, which is indicated by the gray dots.
}
\label{one}
\end{figure}

\begin{figure}
\caption{Spin configurations surrounding a domain wall. (a)
For a $7\times 3$ system with three holes, the width of the lines indicates the 
magnitude of $\langle \vec S_{i}\cdot\vec S_{j}\rangle $ between various sites
when all three holes have been projected onto one of their most
likely configurations.
(b) For the same $7\times 3$ system but with a staggered field applied at
the left end, the length of the arrows indicates $\langle S^z_i\rangle$
when all three holes have been projected onto one of their most
likely configurations.
}
\label{two}
\end{figure}

\begin{figure}
\caption{Hole density and spin moments showing domain walls.
The diameter of the gray holes is proportional to the hole
density $1-\langle n_i\rangle$,
and the length of the arrows is proportional to $\langle S^z_i
\rangle$, according to the scales shown.
(a) A $12\times 3$ system with six holes, 
with a staggered field applied at the left end.
(b) A $16\times 3$ system with eight holes, 
with a staggered field applied along the bottom leg.
}
\label{three}
\end{figure}

\begin{figure}
\caption{Hole rung density for three different densities, on a
$32\times 3$ system.
From bottom to top, the curves show four, twelve, and eighteen
holes, corresponding to $x=0.042$,  $x=0.125$, and  $x=0.1875$.
}
\label{four}
\end{figure}

\begin{figure}
\caption{$n(k) \equiv \langle \psi_0| n_{k_x,k_y,s} |\psi_0\rangle$ for a
three leg ladder and a single chain.
(a) A $32\times 3$ system with $x=0.042$.
(b) A single chain system with $x=0.1$.
In each case, in order to reduce the effects of open boundary 
conditions, an average over many different 
$\langle\psi_0| c^\dagger_{i_x,i_y,s} c_{j_x,j_y,s} |\psi_0\rangle$
with the same separation $(j_x-i_x,j_y-i_y)$ was performed before
Fourier transforming to get $n(k)$. Also, a smooth windowing 
function was applied to remove ``ringing'' near the Fermi
surface.
}
\label{five}
\end{figure}

\begin{figure}
\caption{Same as Fig. 5, for higher densities.
}
\label{six}
\end{figure}

\begin{figure}
\caption{Spin-spin correlations for two different densities on a
$48\times3$ system.
(a) A density of $x=0.042$, showing power-law decay of the 
spin-spin correlations.
(b) A density of $x=0.125$, showing exponential decay of the 
spin-spin correlations.
In each case, many different pairs of points are plotted simultaneously
as a function of their separation,
corresponding to different legs for each point as well as
translations of pairs of points.
}
\label{seven}
\end{figure}

\begin{figure}
\caption{Spin gap for a $44\times 3$ system as a function of doping
$x$.
}
\label{eight}
\end{figure}

\begin{figure}
\caption{Various measurements of pairing.
In (a), the off-diagonal matrix element of $c_{i,\uparrow} c_{j,\downarrow}$ 
between the ground states of a $7\times3$ system with one hole and
three holes is shown. The site $i$ is fixed at the site $(4,2)$, and the
diameter of the gray dots shows the magnitude of the matrix
element as a function of $j$. The sign of the matrix element is
also indicated for each site.
(b) The \dxy\  pairing correlation $D(l)$ is shown for three
different densities, calculated on $32\times3$ ($x=0.1875$) 
and $48\times3$ ($x=0.04$, $x=0.125$) systems.
(c) The same data as in (b) for $x=0.125$ plotted on a semi-log
scale.
}
\label{nine}
\end{figure}

\begin{figure}
\caption{\dxy\  pairing response to a proximity effect pair field operator
for a single chain, and two, three, and four leg ladders. For
the single chain, near-neighbor pairing is measured.
}
\label{ten}

\end{figure}
\begin{figure}
\caption{Typical hole configurations in the transverse domain
wall regime. Results for both (a) and (b) are from 
a $12\times3$ system with six holes, as in Fig. 1(c).
(a) Typical hole configurations for all six holes sampled randomly, using a
classical Monte Carlo procedure with probabilities measured with
DMRG.
(b) Typical hole configurations giving rise to \dxy\  pairing correlations.
The positions of four holes are shown by black dots when
the other two holes are ``fluctuating'' between  the positions
shown by gray dots.
}
\label{eleven}
\end{figure}

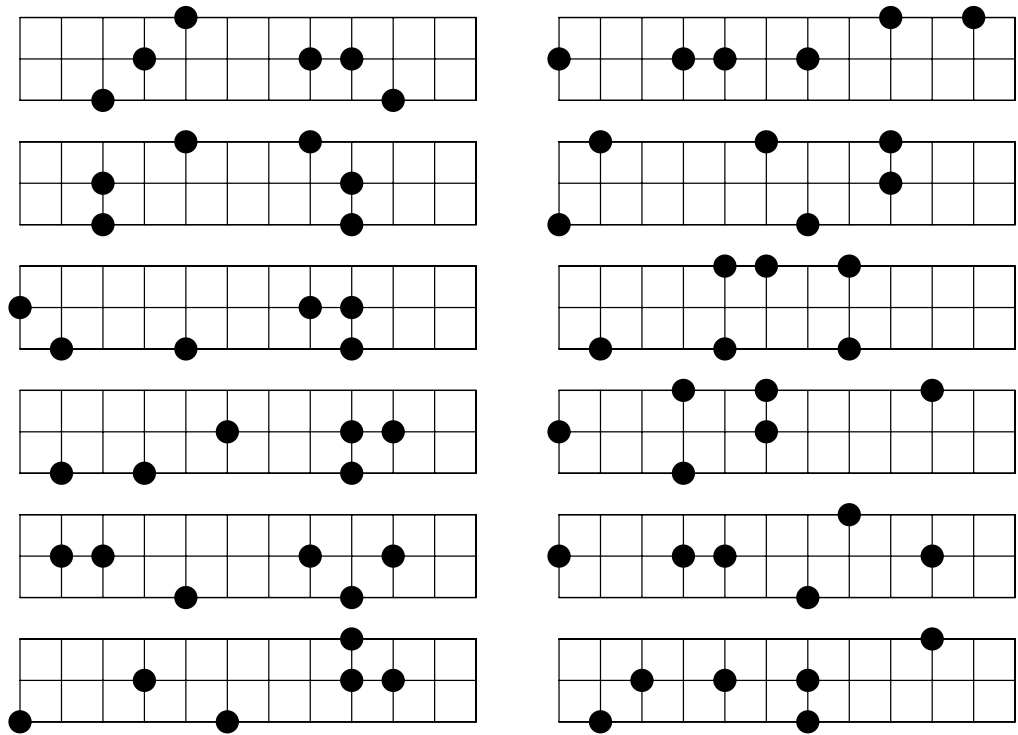
\end{document}